# Dynamic Monte Carlo algorithm for out-of-equilibrium processes in colloidal dispersions

Daniel Corbett,[a] Alejandro Cuetos,[b] Matthew Dennison,[c] and Alessandro Patti[⋆,a]

Colloids have a striking relevance in a wide spectrum of industrial formulations, spanning from personal care products to protective paints. Their behaviour can be easily influenced by extremely weak forces, which disturb their thermodynamic equilibrium and dramatically determine their performance. Motivated by the impact of colloidal dispersions in fundamental science and formulation engineering, we have designed an efficient Dynamic Monte Carlo (DMC) approach to mimic their out-of-equilibrium dynamics. Our recent theory, which provided a rigorous method to reproduce the Brownian motion of colloids by MC simulations, is here generalised to reproduce the Brownian motion of colloidal particles during transitory unsteady states, when their thermodynamic equilibrium is significantly modified. To this end, we investigate monodisperse and bidisperse rod-like particles in the isotropic phase and apply an external field that forces their reorientation along a common direction and induces an isotropic-to-nematic phase transition. We also study the behaviour of the system once the external field is removed. Our simulations are in excellent quantitative agreement with Brownian Dynamics simulations when the DMC results are rescaled with a time-dependent acceptance ratio, which depends on the strength of the applied field. Further generalising our DMC algorithm to processes displaying significant density fluctuations, such as nucleation and growth, where the MC acceptance ratio is expected to depend on both time and space, is currently under investigation.

## 1 Introduction

Colloids are two-phase systems consisting of small objects (molecules, particles, droplets) homogeneously dispersed in a medium. Both dispersed and continuous phase can be either in the gaseous, liquid or solid state. For instance, an emulsion is a colloidal dispersion of liquid droplets in another liquid, while a foam is a dispersion of gas droplets in a liquid or a solid. Perhaps the most common colloidal dispersions are made of solid particles dispersed in a liquid (*e.g.* blood, ink and paints) and are referred to as colloidal suspensions. The dispersed particles should be small enough to remain dispersed in the medium rather than sedimenting. To this end, the thermal energy that keeps them suspended needs to compensate the gravitational potential energy that promotes sedimentation. Equivalently, the particle size should not be larger than the gravitational length defined as [1]:

$$\lambda_{\text{sed}} = \frac{k_B T}{\Delta \rho V_p g} \quad (1)$$

where $k_B \simeq 1.381 \times 10^{-23}$ J K$^{-1}$ is the Boltzmann constant, $T$ the absolute temperature, $\Delta\rho$ the difference between the particle and solvent density, $V_p$ the volume occupied by a particle, and $g$ the gravitational acceleration. At ambient temperature and for a density difference in the order of 100 kg m$^{-3}$, the upper value of the particle radius above which sedimentation would most probably be observed is in the order of $R = 1$ $\mu$m. As far as the lower size is concerned, the IUPAC sets it in the order of 1 nm, specifying that this constraint should be applied to at least one of the particle dimensions[2]. In other words, colloidal particles are supramolecular nanoparticles evenly dispersed in a fluid, whose character-istic length scale is significantly larger than that of conventional molecules[3]. This length scale directly affects the dynamics of colloids and the time scales of their processes. More specifically, colloidal particles show a distinctive diffusive behaviour, usually referred to as Brownian motion[4], that consists of persistent jumpy moves stemming from a kinetic energy contribution that is being dissipated as a result of the collisions with the surrounding molecules of the medium. If the medium is a liquid of viscosity $\mu$, this energy dissipation takes the form of a viscous damping as the Stokes-Einstein equation shows:

$$D = \frac{k_B T}{\varphi} \quad (2)$$

where $D$ is the particle diffusion coefficient and $\varphi = \varphi(\mu)$ the friction coefficient. In case of anisotropic colloidal particles, like rods and disks, rotational diffusion plays a role as crucial as that of translational diffusion. This is especially true when an electric field is applied to charged particles. While this field would force the particles to align along a common direction, their thermal motion, which rather promotes random orientations, would hamper this reorientation. The rotational diffusivity can still be estimated by applying Eq. 2 with a rotational friction coefficient, $\varphi_{\mathrm{rot}}$ that depends on the particle volume.

[a] *School of Chemical Engineering and Analytical Science, The University of Manchester, Manchester, M13 9PL, UK; E-mail:alessandro.patti@manchester.ac.uk*
[b] *Department of Physical, Chemical and Natural Systems, Pablo de Olavide University, 41013 Sevilla, Spain*
[c] *Institut für Theoretische Physik, Technische Universität Berlin, Hardenbergstrasse 36, 10623 Berlin, Germany*

Due to this random drifting, the dynamics of colloids can be investigated by molecular simulation techniques that employ a stochastic rather than deterministic approach, such as Brownian Dynamics (BD) or Dynamic Monte Carlo (DMC), where the presence of the solvent is implicitly incorporated in the effective inter-particle interactions. Deterministic simulation techniques, such as Molecular Dynamics (MD), that make use of Newton dynamics to generate the particle trajectories, are less suitable to describe the Browian motion of colloids. While the BD technique mimics the time-evolution of the particles by integrating stochastic differential equations, DMC simulations follow a statistical approach where particle positions and orientations are updated with a probability satisfying the condition of simple balance[5–7]. Both techniques neglect the effect of the flow field induced by the diffusion of a given particle on all the other particles, commonly referred to as hydrodynamic interactions (HI). Even though their importance has been recognised in some colloidal processes, such as in vivo macromolecular diffusion in cells[8], less clear is their effect in others, such as dense dispersions of colloidal rods[9]. In any case, incorporating them in a simulation is an especially challenging task due to their long-range and many-body nature. A number of more computationally demanding simulation techniques, such as multi-particle collision dynamics[10], lattice Boltzmann[11,12], and stochastic rotation dynamics[13,14], can be employed when HI are expected to be particularly relevant.

We have recently proposed a DMC algorithm to investigate the Brownian motion of pure systems[15] and mixtures[7] of colloidal particles in isotropic, nematic and smectic liquid crystal phases. By rescaling the MC time step with the acceptance ratio of particle displacements and rotations, we demonstrated the existence of a unique MC time scale that allows for a direct comparison with BD simulations. Alternative algorithms have also been proposed by other authors for spherical[16,17], patchy[18] and anisotropic colloidal particles[17,19]. In particular, Sanz and coworkwers have performed a trial-and-error procedure, consisting of few short preliminary simulations, to set the value of the maximum MC displacement, $\delta_t$, and rotation, $\delta_r$, meeting the condition $3\delta_t/\delta_r = \sigma\sqrt{\mathscr{A}_r/\mathscr{A}_t}$, with $\sigma$ the particle diameter, and $\mathscr{A}_t$ and $\mathscr{A}_r$ the acceptance ratio of particle displacements and rotations, respectively[18]. The DMC algorithm that we formulated avoids this prior iterative simulation altogether for monodisperse systems, thus providing an efficient procedure to study the dynamics of colloids. In particular, we showed that our DMC technique could soundly reproduce the BD simulations when the MC time step was sufficiently small to guarantee a uniform acceptance ratio, $\mathscr{A}$, in the $f$-dimensional space set by the maximum variation of the particle's degrees of freedom. Under these conditions, one can assume the acceptance ratio to be independent of the particle's elementary displacements and rotations. This condition, which appears to be rather restrictive, provided excellent results within the typical range of acceptance probabilities of a standard MC simulation. Rather than rescaling with acceptance probability, Jabbari-Farouji and Trizac find an excellent agreement between BD and DMC simulations by equating the short-time self-diffusion extracted from simulations with the infinite-dilution diffusion coefficient[17]. Their method allows to apply DMC to relatively larger elementary moves and thus smaller acceptance ratios, which are, however, very similar to those that we measured in monodisperse systems of rods[7].

In the present work, we extend our theoretical framework to the case in which an external stimulus perturbs the thermodynamic equilibrium of a colloidal system. From a steady-state condition of dynamic equilibrium, where all the observables, including the above mentioned acceptance ratio $\mathscr{A}$, are independent of time ($t$), the system undergoes a transitory unsteady state taking it to a new equilibrium configuration. We apply our DMC simulation technique to simulate the effect of an external field forcing an isotropic phase of rod-like colloidal particles to reorient along a common direction and thus form a nematic liquid crystal. We will show that, even when $\mathscr{A} = \mathscr{A}(t)$, our DMC simulations, which are in excellent quantitative agreement with BD simulations, can be employed to extract reliable dynamical information also from out-of-equilibrium systems. The case of inhomogeneous systems, *e.g.* systems with a density gradient, where the acceptance probability, in the most general case, is also a function of the spatial coordinates, is currently under investigation.

The remainder of this paper is organised as follows. In sec-tion 2, we discuss the theoretical background for the DMC method and consider how to extend the method to the case of an external applied field. In section 3, we introduce the model systems of particles and forces we will investigate using the BD and DMC methods. In section 4, we report and compare the results of our BD and DMC simulations. Finally in section 5, we draw our conclusions.

## 2 Theory

### 2.1 DMC Simulations

In this section, we investigate the link between the evolution of a system of particles in Brownian motion and in MC dynamics. Our aim is to establish a consistent time scale linking DMC simulations to BD simulations that is rigourous in the presence of an external applied field. This has already been established in the absence of external fields both for monodisperse and polydisperse systems[7,15,20]. We first recapitulate the results established previously[7] and then discuss their extension to the case incorporating external fields.

Let's consider a system with a single degree of freedom, where a particle $j$, originally located at $x = 0$, is displaced to a new randomly selected position in the interval $[-\delta x, \delta x]$. The acceptance probability of this move is determined by the Metropolis-Hastings algorithm. If the energy change as a result of the move is denoted by $\Delta \mathscr{E}$, then the move is accepted with probability $\min[1, \exp(-\Delta \mathscr{E}/k_B T)]$ where $\min[X,Y]$ returns the smaller of $X$ and $Y$. The acceptance probability will depend on the size of the displacement, approaching unity as $\delta x \to 0$. If we make the simplifying assumption that the acceptance probability $\mathscr{A}$ is constant over the interval $[-\delta x, \delta x]$, the normalised probability that the move is accepted is $P_{\text{move}} = \mathscr{A}/(2\delta x)$. The mean square displacement, limited to a single MC cycle, then reads:

$$\left\langle x^2 \right\rangle = \int_{-\delta x}^{\delta x} x^2 P_{\text{move}} dx = \frac{\mathscr{A}(\delta x)^2}{3}. \quad (3)$$

It is important now to distinguish between an MC move and an MC cycle. An MC move is an attempt to update the position of a single particle, whereas an MC cycle is $N$ MC moves, with $N$ the number of particles being simulated. Generalising to $\mathscr{C}_{MC}$ cycles we have:

$$\left\langle x^2 \right\rangle = \mathscr{C}_{MC} \frac{\mathscr{A}(\delta x)^2}{3}. \quad (4)$$

We can also generalise to higher dimensions. In 3D space, the number of degrees of freedom for a rigid body increases to six: three translations and three rotations (for a particle with axial symmetry there are five degrees of freedom, three translational and two rotational). We consider a general case of rigid particles with $d$ degrees of freedom. A particle is moved from the origin to a point $\xi = (\xi_1, \xi_2, \ldots, \xi_d)$ belonging to a d-dimensional hyperprism of sides $[-\delta \xi_k, \delta \xi_k]$ with $k = 1, 2, \ldots, d$. Once again, we assume the acceptance probability is uniform over the hyperprism, thus $P_{\text{move}} = \mathscr{A}/V_\Xi$ where $V_\Xi = \Pi_{k=1}^d (2\delta \xi_k)$ is the volume of the hyperprism. The mean-square displacements for a single cycle now take the form:

$$\left\langle \xi_k^2 \right\rangle = \frac{\mathscr{A}(\delta \xi_k^2)}{3}. \quad (5)$$

This result allows us to define a timescale $\delta t_{\text{MC}}$ for a single MC cycle which can be related to the timescale of a BD simulation $\delta t_{\text{BD}}$. According to the Einstein relation, the mean-square displacement of degree of freedom $k$ in a timestep $\delta t_{\text{BD}}$ is given by $\left\langle \xi_k^2 \right\rangle = 2D_k \delta t_{\text{BD}}$, where $D_k$ is the self-diffusion coefficient for degree of freedom $k$. If we define the extent of our hyperprism to be $\delta \xi_k^2 = 2D_k \delta t_{\text{MC}}$ and then equate the BD and MC expressions for $\left\langle \xi_k^2 \right\rangle$, we obtain:

$$\delta t_{\text{BD}} = \frac{\mathscr{A}}{3} \delta t_{\text{MC}}. \quad (6)$$

As shown in [15], the situation is different, but not radically so, if we consider a polydisperse system. In this case, the BD timescale is the same for all populations, but the MC timescale is population dependent. For each population $q$, we have an acceptance ratio $\mathscr{A}_q$ and an MC timescale $\delta t_{q,\text{MC}}$ and they must obey the following relationship:

$$\delta t_{\text{BD}} = \frac{\mathscr{A}_1 \delta t_{1,\text{MC}}}{3} = \frac{\mathscr{A}_2 \delta t_{2,\text{MC}}}{3} = \ldots = \frac{\mathscr{A}_q \delta t_{q,\text{MC}}}{3} \quad (7)$$

For a monodisperse *non-equilibrium* system subject to an external field, the acceptance ratio changes as a function of time. Consequently, we can generalize Eq. 6 as:

$$\delta t_{\text{BD}} = \frac{\mathscr{A}_c}{3} \delta t_{\text{MC}} \quad (8)$$

where we have now made clear that the acceptance ratio $\mathscr{A}_c$, being the acceptance ratio calculated at the $c$th MC cycle, is a function of time. In principle, if we can take the limit as $\delta t_{\text{BD}} \to 0$ and $\delta t_{\text{MC}} \to 0$, we can integrate this expression. In practice, one can only determine $\mathscr{A}_c$ by performing an MC cycle at a fixed $\delta t_{\text{MC}}$ and integrating this expression numerically:

$$t_{\text{BD}}(\mathscr{C}_{\text{MC}}) = \delta t_{\text{MC}} \sum_{c=0}^{\mathscr{C}_{\text{MC}}} \frac{\mathscr{A}_c}{3} \quad (9)$$

where $t_{\text{BD}}(\mathscr{C}_{\text{MC}})$ is the Brownian time after $\mathscr{C}_{\text{MC}}$ MC cycles. Polydisperse *non-equilibrium* systems are slightly different. We have a single BD timescale, for population 1 we keep $\delta t_{q,\text{MC}}$ fixed and calculate the equivalent BD time using Eq. 9. For the remaining populations, we update $\delta t_{q,\text{MC}}$ each cycle according to

$$\delta t_{q,\text{MC}}(\mathscr{C}_{\text{MC}} + 1) = \delta t_{1,\text{MC}}(\mathscr{C}_{\text{MC}}) \frac{\mathscr{A}_1(\mathscr{C}_{\text{MC}})}{\mathscr{A}_q(\mathscr{C}_{\text{MC}})} \quad (10)$$

where the acceptance ratios $\mathscr{A}_1$ and $\mathscr{A}_q$ need to be calculated each cycle.

The fundamental assumption underlying the above results is that we may assume the acceptance ratio is constant over the hyperprism. This is only strictly true if there are no forces acting over the extent of the hyperprism and is thus invalid if external forces are considered. Sanz and Marenduzzo [16] consider a slightly less restrictive scenario where there is a force acting over the hyperprism. The hyperprism is considered sufficiently small that the force can be treated as constant over its volume. For 1D systems, their results are exact:

$$\mathscr{A} = 1 - \frac{\beta|f|\delta x}{4} + \mathcal{O}(\beta|f|\delta x)^2 \tag{11}$$

$$\langle \delta x^2 \rangle = \frac{\delta x^2}{3}\left(1 - 3\frac{\beta|f|\delta x}{8} + \dots\right) \tag{12}$$

$$= \frac{\delta x^2}{3}\left(\frac{3\mathscr{A}}{2} - \frac{1}{2}\right) \tag{13}$$

which is somewhat different to the result obtained in Eq. 3, i.e. the variation of the acceptance ratio over the hyperprism leads to a different rescaling. Exact results are harder to obtain in higher dimensions. For this case, we consider a Gaussian Approximation (GA) (see appendix A for details), which becomes increasingly accurate as we increase the number of degrees of freedom. The acceptance ratio may be approximated by:

$$\mathscr{A} = \frac{1}{V_\Xi}\int_{V_\Xi}\min(1,\exp(-\beta\mathbf{f}\cdot\boldsymbol{\xi}))dV$$

$$\approx 1 - 0.23\sqrt{\sum_i(\beta f_i \delta\xi_i)^2} \tag{14}$$

where $\mathbf{f} = (f_1, f_2, \dots, f_d)$ is a constant force acting over the hyperprism. We note that this approximation is very close to the analytic result for 1D (0.23 vs 0.25). Similarly, one can calculate an approximation for $\langle \xi_k^2 \rangle$:

$$\langle \xi_k^2 \rangle = \frac{\delta\xi_k^2}{3}\underbrace{\left[1 - 0.23\sqrt{0.8(\beta f_k \delta\xi_k)^2 + \sum_i(\beta f_i \delta\xi_i)^2}\right]}_{\mathscr{A}_2} \tag{15}$$

The term we have labelled $\mathscr{A}_2$ is clearly not equal to $\mathscr{A}$. However, as the number of degrees of freedom increases, the importance of the $0.8(\beta f_k \delta\xi_k)^2$ vs the $\sum_i(\beta f_i \delta\xi_i)^2$ term should diminish, i.e. we expect $\mathscr{A}$ to be an increasingly good approximation to $\mathscr{A}_2$ as the number of degrees of freedom increases. To demonstrate this, consider the case of a force of magnitude $k_B T/\sigma$ pointing in a random direction. For the sake of simplicity we will assume the hyperprism is a hypercube i.e. $\delta\xi_k = \delta\xi$ for all k. Results for $(1 - \mathscr{A}_2)/(1 - \mathscr{A})$ calculated exactly (d=1...4) and using the GA (d=1...6) are shown in table 1. It is clear that as the number of degrees of freedom increases $\mathscr{A}_2 \to \mathscr{A}$.

| d | $(1-\mathscr{A}_2)/(1-\mathscr{A})$ (Exact) | $(1-\mathscr{A}_2)/(1-\mathscr{A})$ (GA) |
|---|---|---|
| 1 | 1.5 | 1.34 |
| 2 | 1.23 | 1.18 |
| 3 | 1.15 | 1.12 |
| 4 | 1.10 | 1.09 |
| 5 | - | 1.07 |
| 6 | - | 1.06 |

**Table 1** The ratio $(1 - \mathscr{A}_2)/(1 - \mathscr{A})$ as a function of the number of degrees of freedom for a force of magnitude $k_B T/\sigma$ pointing in a random direction. We have assumed the hyperprism associated with the particle is a hypercube of extent $\pm\delta\xi$ for each degree of freedom. For the exact results, we calculate $10^5$ vectors isotropically distributed over a d-dimensional hypersphere and then calculate $\mathscr{A}$ and $\mathscr{A}_2$ by numerical integration for each vector. This calculation becomes increasingly slow as the number of degrees of freedom increases. However, we see the GA results provide an increasingly good approximation to the exact results as $d$ increases.

To summarize, we propose that to rescale an MC cycle into real time for *out-of-equilibrium* systems we should use the scaling shown in Eq. 9. Additionally, we expect the accuracy of this rescaling to increase as the number of degrees of freedom of the particles increases. In the following sections, we will apply our rescaling to spherocylinders, which have 5 degrees of freedom.

## 3 Model and Methodology

We investigate the behaviour of two colloidal systems, one a collection of monodisperse spherocylinders with length-to-diameter ratio $L/\sigma = 5$, and the other a racemic bidisperse mixture of spherocylinders with $L/\sigma = 3$ and $L/\sigma = 7$. The particles interact with each other via a purely repulsive shifted and truncated Kihara potential of the form [21,22]:

$$U_{ij} = \begin{cases} 4\varepsilon\left[\left(\frac{\sigma}{d_m}\right)^{12} - \left(\frac{\sigma}{d_m}\right)^6 + \frac{1}{4}\right] & d_m \leq 2^{1/6}\sigma \\ 0, & d_m > 2^{1/6}\sigma \end{cases} \tag{16}$$

where $U_{ij} = U_{ij}(\mathbf{r}_{ij}, \hat{\mathbf{u}}_i, \hat{\mathbf{u}}_j)$. The subscripts $i$ and $j$ refer to a pair of interacting spherocylinders, $\mathbf{r}_{ij}$ is the vector connecting their centers, $\hat{\mathbf{u}}_i$ and $\hat{\mathbf{u}}_j$ are unit vectors describing their orientations, $\varepsilon$ the strength of their interaction, and $d_m = d_m(\mathbf{r}_{ij}, \hat{\mathbf{u}}_i, \hat{\mathbf{u}}_j)$ is the minimum distance between them [23]. We use $\sigma$, $\varepsilon$ and $\tau = \sigma^2/D_0$ as our length, energy and time units, with $D_0 = k_B T/(\mu\sigma)$ a diffusion constant and $\mu$ the shear viscosity of the solvent. We have used $T^* = k_B T/\varepsilon = 1.465$ because at this temperature the phase behaviour of soft spherocylinders resembles that of hard spherocylinders of the same length and diameter [21,22]. In addition to interacting with each other, the particles are subject to an external field which couples to the long axis $\hat{\mathbf{u}}$ of the spherocylinder. We will assume the following form for the external potential $U^{\text{ext}}$:

$$U^{\text{ext}} = -\frac{3\lambda}{2} \sum_{i=1}^{N} (\hat{\mathbf{u}}_i \cdot \hat{\mathbf{e}})^2 \tag{17}$$

where $\lambda$ represents the strength of the applied field and $\hat{\mathbf{e}}$ is the direction of the applied field. This external field could for example correspond to an electric field coupling to the dielectric anisotropy of the spherocylinders.

The parameter whose time evolution we will measure is the scalar nematic order parameter $\langle P_2 \rangle$, which is the largest eigenvalue of the order parameter tensor defined by:

$$\mathbf{S}_{\alpha,\beta} = \frac{1}{2N} \sum_{i=1}^{N} (3\hat{\mathbf{u}}_{\alpha,i}\hat{\mathbf{u}}_{\beta,i} - \delta_{\alpha,\beta}). \tag{18}$$

where $\delta_{\alpha,\beta}$ is the Kronecker delta. For the bidisperse case, we can monitor the scalar nematic order parameter for each population separately.

For both monodisperse and bidisperse cases we perform two types of simulations, denoted ON and OFF. For an ON simulation, we start with a single initial isotropic configuration at a specified density $\rho$. A field of strength $\lambda$ is applied and the subsequent evolution of the system is monitored up to $t_{\text{BD}}/\tau = 300$. For each simulation, we average over 100 trajectories, which start from the same initial configuration, but have a different seed for the random number generator. By contrast, for an OFF simulation, we start with a single configuration at a specified density $\rho$, which has equilibrated in a field of strength $\lambda$. The field is removed at $t_{\text{BD}} = 0$ and the subsequent evolution of the system is monitored up to $t_{\text{BD}}/\tau = 300$. Once again, we average over 100 values of the seed for the random number generator.

### 3.1 DMC Simulations

We perform DMC simulations in the NVT ensemble, with simultaneous attempts to displace and rotate a selected particle. Translational and rotational moves are accepted/rejected according to the standard Metropolis-Hasting algorithm, that is with probability $\min[1, \exp(-\Delta\mathscr{E}/k_B T)]$ where $\Delta\mathscr{E}$ is the change in energy caused by the proposed move. If the move is accepted, the system energy is increased by $\Delta\mathscr{E}$. The energy $\Delta\mathscr{E}$ contains contributions from the mutual interactions between the particles and from the interaction between the selected partcle and the external field. Since our aim here is to mimic physical systems, we do not allow unphysical moves such as swaps or cluster moves. For the displacement of the center of mass of a given particle $j$, the motion is decoupled into three terms $\delta \mathbf{r}_j = X_{\parallel}\hat{\mathbf{u}}_j + X_{\perp,1}\hat{\mathbf{v}}_{j,1} + X_{\perp,2}\hat{\mathbf{v}}_{j,2}$, where $\hat{\mathbf{u}}_j$ is a unit vector parallel to the long axis of the spherocylinder and $\hat{\mathbf{v}}_{j,m}$ are two randomly chosen vectors that are perpendicular to each other and to $\hat{\mathbf{u}}_j$. The magnitude of the displacements are selected randomly from uniform distibutions satisying $|X_{\parallel}| \leq \sqrt{2D_{\parallel,j}\delta t_{\text{MC}}}$ and $|X_{\perp,m}| \leq \sqrt{2D_{\perp,j}\delta t_{\text{MC}}}$ where $D_{\parallel,j}$ and $D_{\perp,j}$ are the self-diffusion coefficients of the spherocylinder along and perpendicular to $\hat{\mathbf{u}}_j$. For rotations, the vector $\hat{\mathbf{u}}_j$ changes to $\hat{\mathbf{u}}_j + \delta\hat{\mathbf{u}}_j$ with $\delta\hat{\mathbf{u}}_j = Y_{\theta,1}\hat{\mathbf{w}}_{j,1} + Y_{\theta,2}\hat{\mathbf{w}}_{j,2}$. The two vectors $\hat{\mathbf{w}}_{j,m}$ are randomly chosen and are perpendicular both to each other and to $\hat{\mathbf{u}}_j$. The magnitudes of the rotations are selected randomly from a uniform distributions satisfying $|Y_{\theta,m}| \leq \sqrt{2D_{\theta,j}\delta t_{\text{MC}}}$ where $D_{\theta,j}$ is the rotational self-diffusion coefficient of the spherocylinder.

For the diffusion coefficients, we use the analytic results available for prolate ellipsods [24]:

$$D_{\perp,j} = D_0 \frac{(2a^2-3)K + 2a}{16\pi(a^2-1)}, \tag{19}$$

$$D_{\parallel,j} = D_0 \frac{(2a^2-1)K - 2a}{8\pi(a^2-1)}, \tag{20}$$

$$D_{\theta,j} = 12\frac{D_0}{\sigma^2} \frac{(2a^2-1)K - 2a}{16\pi(a^4-1)} \tag{21}$$

where $a = L/\sigma$, $b = \sigma/2$, and $K$ is given by:

$$K = \frac{2}{\sqrt{a^2-1}} \log\left[a + \sqrt{a^2-1}\right] \tag{22}$$

We describe our algorithm in more detail for the bidisperse case. Let's suppose that we have two populations of particles with $N_1$ of type 1 and $N_2$ of type 2 and $N = N_1 + N_2$ in total. We have two MC timesteps, $\delta t_{1,\text{MC}}$ and $\delta t_{2,\text{MC}}$. We start with $\delta t_{1,\text{MC}} = \delta t_{2,\text{MC}}$, and calculate the extent of each degree of freedom of our hyperprisms according to $\delta \xi_{k,q} = \sqrt{2D_{k,q}\delta t_{\text{MC},q}}$ where $D_{k,q}$ is the self-diffusion coefficient for degree of freedom $k$ for population $q$. Starting with $t_{\text{BD}} = 0$, we then:

1. Pick a particle at random, perform a random translation and rotation and then determine if the move is accepted or not via the Metropolis-Hastings method.

2. Repeat this process $N$ times.

3. Calculate the acceptance rates $\mathscr{A}_1$ and $\mathscr{A}_2$ for the two populations.

4. Keep $\delta t_{1,\text{MC}}$ fixed, scale $\delta t_{2,\text{MC}}$ according to $\delta t_{2,\text{MC}} = \mathscr{A}_1 \delta t_{1,\text{MC}}/\mathscr{A}_2$.

5. Update the dimensions of the hyperprism of population 2 according to the new $\delta t_{2,\text{MC}}$.

6. Update the BD time according to $t_{\text{BD}} = t_{\text{BD}} + \delta t_{1,\text{MC}}\mathscr{A}_1/3$.

7. Iterate the process to advance $t_{\text{BD}}$.

### 3.2 BD Simulations

In a BD simulation, a stochastic differential equation, the so-called Langevin equation, is integrated forward in time and trajectories of particles are created [25,26]. As before, let $\mathbf{r}_j$ be the position of the center of mass of rod $j$, with $\hat{\mathbf{u}}_j$ the unit vector oriented along the long axis of $j$, $\hat{\mathbf{v}}_{j,m}$ and $\hat{\mathbf{w}}_{j,m}$, with $m=1$ or 2, two independent pairs of perpendicular unit vectors being also perpendicular to $\hat{\mathbf{u}}_j$. Furthermore, $\mathbf{F}_j$ and $\mathbf{T}_j$ are the total force and torque acting over the particle $j$. To compute them for systems of particles interacting via a Kihara potential, we refer to the work by Vega and Lago[27]. Over a BD time step, being set to $\Delta t = 10^{-5}\tau$, the position of the center of mass and the orientation of each particle are updated as follows

$$\mathbf{r}_j^{\parallel}(t+\Delta t) = \mathbf{r}_j^{\parallel}(t) + \frac{D_{\parallel,j}}{k_B T}\mathbf{F}_j^{\parallel}(t)\Delta t + (2D_{\parallel,j}\Delta t)^{1/2} R_{\parallel}\hat{\mathbf{u}}_j(t) \quad (23)$$

$$\mathbf{r}_j^{\perp}(t+\Delta t) = \mathbf{r}_j^{\perp}(t) + \frac{D_{\perp,j}}{k_B T}\mathbf{F}_j^{\perp}(t)\Delta t$$
$$+ (2D_{\perp,j}\Delta t)^{1/2}(R_1^{\perp}\hat{\mathbf{v}}_{j,1}(t) + R_2^{\perp}\hat{\mathbf{v}}_{j,2}(t)) \quad (24)$$

$$\hat{\mathbf{u}}_j(t+\Delta t) = \hat{\mathbf{u}}_j(t) + \frac{D_{\vartheta}}{k_B T}\mathbf{T}_j(t)\times\hat{\mathbf{u}}_j(t)\Delta t$$
$$+ (2D_{\vartheta}\Delta t)^{1/2}(R_1^{\vartheta}\hat{\mathbf{w}}_{j,1}(t) + R_2^{\vartheta}\hat{\mathbf{w}}_{j,2}(t)) \quad (25)$$

where $\mathbf{r}_j^{\parallel}$ and $\mathbf{r}_j^{\perp}$ indicate the projections of the positions of particle $j$ along $\hat{\mathbf{u}}_j$ and along the directions perpendicular to $\hat{\mathbf{u}}_j$, respectively; $R^{\parallel}, R_1^{\perp}, R_2^{\perp}, R_1^{\vartheta}$ and $R_2^{\vartheta}$ are independent Gaussian random numbers of variance 1 and zero mean; and $\mathbf{F}_j^{\parallel}$ and $\mathbf{F}_j^{\perp}$ are the parallel and perpendicular components of the forces, respectively[23,27]. Additionally, the interaction with the external field generates a torque, which is calculated as

$$\mathbf{T}_j^{\text{ext}} = -\hat{\mathbf{u}}_\mathbf{j} \times \bar{\nabla}_{\hat{\mathbf{u}}_\mathbf{j}} U^{\text{ext}}$$
$$= 3\lambda(\hat{\mathbf{u}}_\mathbf{j}\cdot\hat{\mathbf{e}})(\hat{\mathbf{u}}_\mathbf{j}\times\hat{\mathbf{e}}) \quad (26)$$

## 4 Results

### 4.1 Monodisperse Rods

We consider a collection of monodisperse spherocylinders with a length to diameter ratio $L/\sigma = 5$. We perform NVT simulations with 1000 particles in a cubic box at two number densities, $\rho \equiv N/V = 0.008$ and $\rho = 0.023$, corresponding to packing fractions of $\eta \equiv Nv_0/V = 0.036$ and $\eta = 0.104$, respectively, with $v_0$ the volume of a spherocylinder. For each density, we perform ON and OFF simulations for two field strengths $\lambda = 2$ and $\lambda = 5$. In figure 1, we show the ON and OFF simulations for $\eta = 0.036$ and $\lambda = 2$. Solid lines show the results of DMC simulations, while dashed lines correspond to BD simulations. It is possible to observe that, as soon as the field is switched on, the order of the system increases as a result of the reorientation of the particles, which align along a common director and form a nematic phase. The order parameter

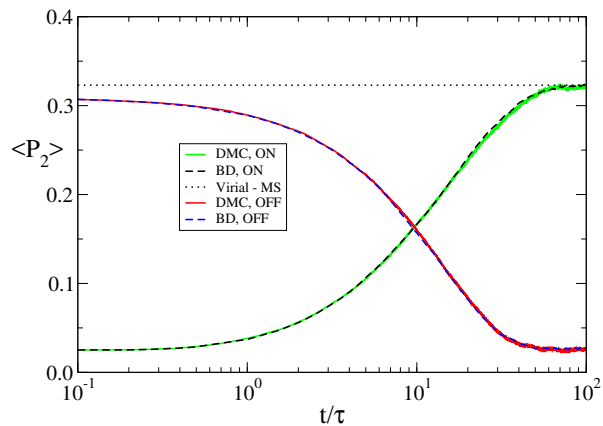

**Fig. 1** Development of the order parameter for a monodisperse system of spherocylinders with $L/\sigma = 5$ and $\eta = 0.036$ when an external field with strength $\lambda = 2$ is applied along the $(1,0,0)$ direction (ON) and subesquently removed (OFF). Solid lines correspond to DMC simulations, dashed lines correspond to BD simulations. We also show the *equilibrium* value of $\langle P_2 \rangle$ for $\lambda = 2$ obtained by calculating the virial coefficients (see appendix B)

increases up to a plateau in approximately 3 time decades, with a value that in principle can depend on system density, particle geometry and field strength. However, we will see that some of these three contributions might not be especially relevant.

Similar considerations are still valid for the OFF experiments, which promote a random reorientation of the particles and thus induce a nematic-to-isotropic phase transition. We observe that this transition to the new equilibrium state also takes roughly 3 time decades, and the plateau value of the order parameter after switching the field off is the same as that measured in the undisturbed isotropic phase. The dynamical evolution of the order parameter is pictured, with the same degree of detail, by DMC and BD simulations, whose quantitative agreement is excellent for both the ON and OFF simulations.

This behaviour is mirrored in figure 2, which shows results obtained when a stronger field is applied. More specifically, at $\lambda = 5$, the final value of the order parameter is approximately twice as large as that measured at $\lambda = 2$, but the time needed to reach the new equilibrium state is still 3 time decades. Consequently, the field strength, at least within the range explored in this study, has a negligible impact on the reorientation time, a feature detected by both BD and DMC simulations. Finally, in figure 3, we show BD and DMC results for a somewhat denser, but still isotropic, system, with packing fraction $\eta = 0.104$, and a field strength of $\lambda = 5$. The dynamical behaviour of the order parameter is quantitatively and qualitatively similar to that already detected and, once again, the agreement between BD and DMC simulations, for

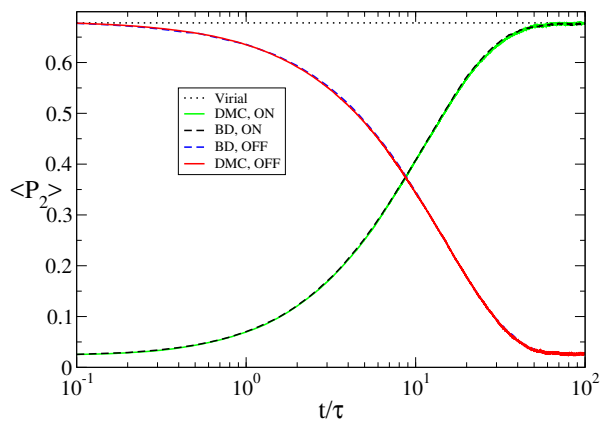

**Fig. 2** Development of the order parameter for a monodisperse system of spherocylinders with $L/\sigma = 5$ and $\eta = 0.036$ when an external field with strength $\lambda = 5$ is applied along the $(1,0,0)$ direction (ON) and subsequently removed (OFF). Solid lines correspond to DMC simulations, dashed lines correspond to BD simulations. We also show the *equilibrium* value of $\langle P_2 \rangle$ for $\lambda = 2$ obtained by calculating the virial coefficients (see appendix B)

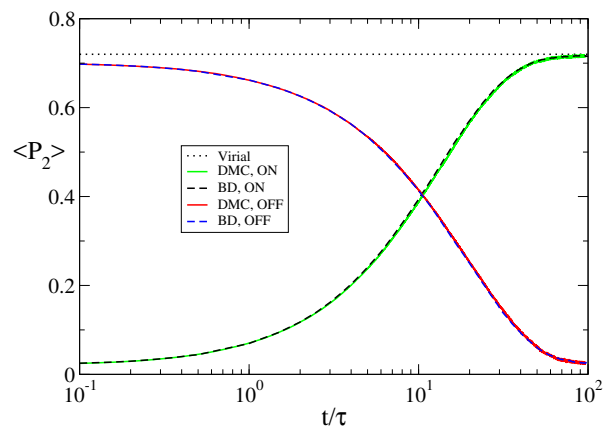

**Fig. 3** Development of the order parameter for a monodisperse system of spherocylinders with $L/\sigma = 5$ and $\eta = 0.104$ when an external field with strength $\lambda = 5$ is applied along the $(1,0,0)$ direction (ON) and subsequently removed (OFF). Solid lines correspond to DMC simulations, dashed lines correspond to BD simulations. We also show the *equilibrium* value of $\langle P_2 \rangle$ for $\lambda = 2$ obtained by calculating the virial coefficients (see appendix B)

both ON and OFF simulations, is excellent.

For all monodisperse cases, we have also estimated the *equilibrium* value of $\langle P_2 \rangle$ as a function of applied field strength ($\lambda$) by calculating the virial coefficients (see appendix B for details). As can be seen in figures 1, 2 and 3, both the DMC and BD methods reach the equilibrium values for the order parameter predicted by theory.

### 4.2 Bidisperse Rods

It is important to understand how the simultaneous presence of more than one species in the system can influence the dependence of the acceptance rates (one for each species) on time. To this end, we now turn our attention to a bidisperse system of 1500 particles in a cubic box with volume $V = (35\sigma)^3$, giving a number density $\rho = 0.035$. Half of the rods have a length $L = 3\sigma$, whereas the length of the remaining half is $L = 7\sigma$. The total packing fraction is given by $\eta = 0.156$, corresponding to an isotropic phase at the thermodynamic equilibrium. As already done for the monodisperse systems, we monitor the growth of the nematic order parameter over time by running BD and DMC simulations. However, this control is performed for each particle population separately. In figure 4, we show the dynamical evolution of the order parameters as functions of time for BD (dashed lines) and DMC (solid lines) for both ON (left frame) and OFF (right frame) simulations for a field of strength $\lambda = 2$. The particle geometry has an evident effect on the kinetics of reorientation and phase transformation, being significantly slower at increas-ing anisotropy. In particular, short particles reorient faster than long particles, but their ordering is slightly weaker, as revealed by the plateau of $\langle P_2 \rangle$ in the new steady state at the end of the ON simulations. Similar to what we already observed for monodisperse systems, the agreement between BD and DMC simulations of out-of-equilibrium bidisperse systems is excellent. This result does not significantly change when the field strength is increased to $\lambda = 5$, as figure 5 indicates. Also in this case, our DMC method is able to reproduce the BD simulation results with a very high degree of accuracy.

### 4.3 Benchmarking

Having established that our DMC simulation technique is able to successfully reproduce the Brownian motion of out-of-equilibrium colloidal suspensions and shows excellent agreement with BD for the dynamical behaviour of $\langle P_2 \rangle$ as a function of applied external field, it is useful to compare the computational time taken for each technique to run. To this end, we measured the CPU time taken to generate a single trajectory with the BD and DMC methods for the ON simulation of a bidisperse system with $\lambda = 5$ and up to a time $t_{BD}/\tau = 10$. For both simulations, we compiled our code with the GNU Fortran compiler with maximum optimisation. Both compiled programs were executed on the same desktop PC running Ubuntu 17.04 and equipped with four 3.60 GHz Intel i7-4790 CPUs and 16.0 GB of RAM. For DMC we require a CPU time of 25 seconds, while for BD the simulation takes 7 minutes, i.e. DMC is 17 times quicker than BD. The efficiency of the DMC method compared to the BD method is not primarily algorithmic - for the same timestep the two methods take approximately the same CPU time to execute. Rather the formulation of the DMC method allows one to apply much larger timesteps than can be used with BD.

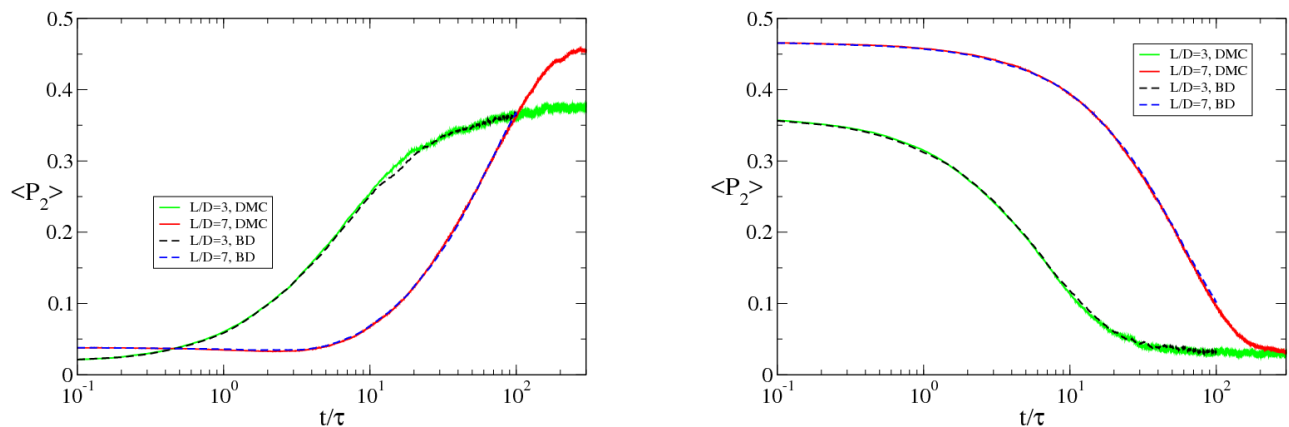

**Fig. 4** Development of the order parameters for a bidisperse system of spherocylinders consisting of a racemic mixture of spherocylinders with $L/\sigma = 3$ and $L/\sigma = 7$ after an external field with strength $\lambda = 2$ is turned on (LHS) and subsequently removed (RHS). The packing fraction is $\eta = 0.156$. Solid lines show DMC results, dashed BD.

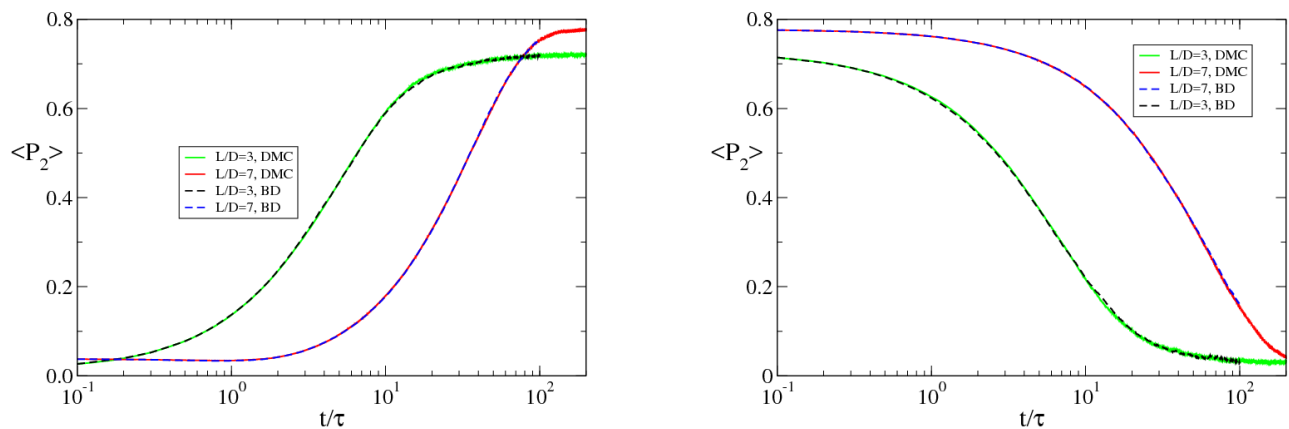

**Fig. 5** Development of the order parameters for a bidisperse system of spherocylinders consisting of a racemic mixture of spherocylinders with $L/\sigma = 3$ and $L/\sigma = 7$ after an external field with strength $\lambda = 5$ is turned on (LHS) and subsequently removed (RHS). The packing fraction is $\eta = 0.156$. Solid lines show DMC results, dashed BD.

## 5 Conclusions

In summary, we have generalised our DMC method to include the study of the out-of-equilibrium dynamics of colloidal suspensions. More specifically, we investigated the reorientation of monodisperse and bidisperse systems of spherocylinders, which are driven out of equilibrium by the application of an external field coupled to their main axis. This reorientation produces an isotropic-to-nematic phase transition when the field is switched on, and the opposite transformation pathway when it is switched off. We have tested the validity of our method by performing a comparative analysis between the DMC method and the BD method. The dynamical property we monitor is the nematic order parameter $\langle P_2 \rangle$, which measures the ordering of the particles along a common direction. For DMC simulations we have monitored this parameter as a function of MC cycles and then rescaled the results onto real (BD) time using the rescaling in Eq. 6. The resulting rescaled curves are shown to overlap very well with the BD results for all the cases we have considered, which include different field strengths, particle geometries and system densities. We have also performed benchmarking of the two methods and find the DMC method to be approximately 17 times faster than the BD method, representing a substantial reduction in computational effort.

The extended DMC method proposed in this work is applicable to any kind of system containing particles with orientational degrees of freedom such as patchy colloids or general anisotropic particles in the colloidal regime. As we have seen, the method works for both monodisperse and polydisperse systems. Nevertheless, the algorithm still comes with some restrictions, perhaps the most important of which is the spatial homogeneity of the driving force. Extending the method to a spatially varying driving force and inhomogeneous systems, where a density gradient is observed, is currently under investigation.

## 6 Conflicts of interest

There are no conflicts to declare.

## 7 Acknowledgments

DC and AP acknowledge financial support from EPSRC under grant agreement EP/N02690X/1. AC acknowledges scholarship PPI1719 from Universidad Pablo Olavide for funding his research visit to School of Chemical Engineering and Analytical Science, The University of Manchester. AC also acknowledges project CTQ2012-32345 funded by the Junta de Andalucía-FEDER and C3UPO for HPC facilities provided. The use of the Computational Shared Facility at the University of Manchester is also acknowledged. DC would like to thank Andrew Masters for useful discussions.

## A Rescaling time. The case of external field

We have derived our time-rescaling in Eq. 6 by assuming a uniform acceptance ratio over the hyper-prism. If external forces/torques are present, this assumption is no longer valid. A constant acceptance ratio would always lead to $\langle \xi_k \rangle = 0$ yet applying an external force/torque should result in a non-zero value for $\langle \xi_k \rangle$. To clarify this point, we first show a simplified explanation and then a more meticulous discussion.

Let's consider a 1D system with a external field applied along the positive direction. The particles are allowed to move at random in the interval $[-\delta x, \delta x]$, with acceptance probability $\mathscr{A}_+$ if the move is in the interval $[0, \delta x]$ and $\mathscr{A}_-$ if in the interval $[-\delta x, 0]$. Consequently, the average displacement reads

$$\langle x \rangle = \int_0^{\delta x} x \frac{\mathscr{A}_+}{2\delta x} dx + \int_{-\delta x}^{0} x \frac{\mathscr{A}_-}{2\delta x} dx = \frac{\delta x}{4}(\mathscr{A}_+ - \mathscr{A}_-) \quad (27)$$

The result is zero if $\mathscr{A}_+ = \mathscr{A}_-$, but otherwise positive or negative depending on the magnitude of each acceptance or, equivalently, on the direction of the field. By contrast, the mean square displacement reads

$$\langle x^2 \rangle = \int_0^{\delta x} x^2 \frac{\mathscr{A}_+}{2\delta x} dx + \int_{-\delta x}^{0} x^2 \frac{\mathscr{A}_-}{2\delta x} dx = \frac{(\delta x)^2}{6}(\mathscr{A}_+ + \mathscr{A}_-) \quad (28)$$

If we define an average acceptance ratio as $\mathscr{A} = (\mathscr{A}_+ + \mathscr{A}_-)/2$, the original result is recovered:

$$\langle x^2 \rangle = \frac{(\delta x)^2}{3}\mathscr{A} \quad (29)$$

We now show that a more rigorous approach can lead to the same conclusions. Above, we have mentioned that, assuming a uniform acceptance ratio, the application of an external force/torque should result in a non-zero value for $\langle \xi_k \rangle$. Following Ref.[16], we relax this condition slightly and, rather than a uniform acceptance ratio, we assume that the extent of our hyper-prism is sufficiently small that we may assume that any forces/torques (both internal and external) acting on a particle are uniform within the hyperprism. The acceptance is then given by:

$$\mathscr{A} = \frac{1}{V_\Xi} \int_{V_\Xi} \min(1, \exp(-\beta \mathbf{f} \cdot \xi)) dV$$

$$= \frac{1}{V_\Xi} \int_{V_\Xi} [1 + \min(0, \exp(-\beta \mathbf{f} \cdot \xi) - 1)] dV \quad (30)$$

where $V_\Xi$ is the volume of the hyperprism. We will further assume that for each dimension of the hyperprism we have $\beta f_i \delta \xi_i << 1$ which allows us to expand the exponential, therefore:

$$\mathscr{A} = \frac{1}{V_\Xi} \int_{V_\Xi} [1 + \min(0, -\beta \mathbf{f} \cdot \xi)] dV \quad (31)$$

$$= \frac{1}{V_\Xi} \int_{V_\Xi} [1 - \beta \mathbf{f} \cdot \xi H(\beta \mathbf{f} \cdot \xi)] dV \quad (32)$$

where $H(x)$ is the Heaviside step function. In general evaluation of this integral is difficult. It is convenient to introduce an integral representation for the Heaviside function as follows:

$$H(x) = \lim_{\varepsilon \to 0^+} \frac{1}{2\pi i} \int_{-\infty}^{\infty} \frac{1}{\tau - i\varepsilon} \exp(ix\tau) d\tau \quad (33)$$

where $\varepsilon$ approaches 0 from above. Using this integral representa-tion we can calculate the acceptance rate:

$$\mathscr{A} = 1 - \frac{1}{2\pi i V_\Xi} \int\int \frac{1}{\tau - i\varepsilon} \beta \mathbf{f} \cdot \boldsymbol{\xi} \exp(i\beta \mathbf{f} \cdot \boldsymbol{\xi} \tau) d\tau \quad (34)$$

$$= 1 + \frac{1}{2\pi V_\Xi} \int_{\tau=-\infty}^{\infty} \frac{1}{\tau - i\varepsilon} \frac{\partial}{\partial \tau} \left[ \int \exp(i\beta \mathbf{f} \cdot \boldsymbol{\xi} \tau) dV \right] d\tau \quad (35)$$

$$= 1 + \frac{1}{2\pi V_\Xi} \int_{\tau=-\infty}^{\infty} \frac{1}{\tau - i\varepsilon} \frac{\partial}{\partial \tau} \left[ \prod_{i=1}^{N} 2 \frac{\sin(\beta f_i \delta \xi_i \tau)}{\beta f_i \tau} \right] d\tau$$

$$= 1 + \frac{1}{2\pi} \int_{\tau=-\infty}^{\infty} \frac{1}{\tau - i\varepsilon} \frac{\partial}{\partial \tau} \left[ \prod_{i=1}^{N} \frac{\sin(\beta f_i \delta \xi_i \tau)}{\beta f_i \delta \xi_i \tau} \right] d\tau \quad (36)$$

For a particular force we can calculate this integral by numerical integration. However, we can also make some analytic progress by use of a Gaussian Approximation (GA). We consider the power series expansion of sinc($\alpha x$)

$$\text{sinc}(\alpha x) = 1 - \frac{(\alpha x)^2}{6} + \mathcal{O}((\alpha x)^4) \quad (37)$$

$$\approx \exp(-\alpha^2 x^2/6). \quad (38)$$

Using the GA, we can approximate Eq. 36 as:

$$\mathscr{A} = 1 + \frac{1}{2\pi} \int_{\tau=-\infty}^{\infty} \frac{1}{\tau - i\varepsilon} \frac{\partial}{\partial \tau} \left[ \prod_{i=1}^{N} \frac{\sin(\beta f_i \delta \xi_i \tau)}{\beta f_i \delta \xi_i \tau} \right] d\tau$$

$$\approx 1 + \frac{1}{2\pi} \int_{\tau=-\infty}^{\infty} \frac{1}{\tau - i\varepsilon} \frac{\partial}{\partial \tau} \left[ \exp\left( -\frac{\sum_i (\beta f_i \delta \xi_i)^2 \tau^2}{6} \right) \right] d\tau$$

$$\approx 1 - \sqrt{\frac{1}{6\pi}} \sqrt{\sum_i (\beta f_i \delta \xi_i)^2}$$

$$\approx 1 - 0.23 \sqrt{\sum_i (\beta f_i \delta \xi_i)^2} \quad (39)$$

Using the GA we are also able to calculate $\langle \xi_k^2 \rangle$ using:

$$\langle \xi_k^2 \rangle = \frac{\delta \xi_k^2}{3} \underbrace{\left[ 1 - 0.23 \sqrt{0.8(\beta f_k \delta \xi_k)^2 + \sum_i (\beta f_i \delta \xi_i)^2} \right]}_{\mathscr{A}_2}. \quad (40)$$

Comparing the term labelled $\mathscr{A}_2$ with Eq. 39, we see that $\mathscr{A}_2$ differs from $\mathscr{A}$. However, as the number of degrees of freedom increases, the importance of the $0.8(\beta f_k \delta \xi_k)^2$ term vs the $\sum_i (\beta f_i \delta \xi_i)^2$ term should diminish, i.e. we expect $\mathscr{A}$ to be an increasingly good approximation to $\mathscr{A}_2$ as the number of degrees of freedom of the particles increases. As an example, consider the case where we have $\beta f_p \delta \xi_p \equiv \gamma \ll 1$ for each degree of freedom.

We then have:

$$\mathscr{A}_2 = 1 - 0.23 \sqrt{0.8(\beta f_k \delta \xi_k)^2 + \sum_i (\beta f_i \delta \xi_i)^2}$$

$$= 1 - 0.23 \gamma \sqrt{d} \sqrt{1 + \frac{0.8(\beta f_k \delta \xi_k)^2}{\sum_i (\beta f_i \delta \xi_i)^2}}$$

$$= 1 - 0.23 \gamma \sqrt{d} \left( 1 + \frac{0.4}{d} - \frac{1}{8}\left(\frac{0.8}{d}\right)^2 + \ldots \right)$$

$$= \mathscr{A} - 0.23 \gamma \left( \frac{0.4}{\sqrt{d}} - \frac{1}{8} \frac{0.8^2}{d^{3/2}} + \ldots \right)$$

$$= \mathscr{A} - (1 - \mathscr{A}) \left( \frac{0.4}{d} - \frac{1}{8} \frac{0.8^2}{d^2} + \ldots \right) \quad (41)$$

and therefore:

$$\frac{1 - \mathscr{A}_2}{1 - \mathscr{A}} = 1 + \frac{0.4}{d} - \frac{0.08}{d^2} \quad (42)$$

i.e. we see that $\mathscr{A}_2 \to \mathscr{A}$ as $d$ increases.

## B Equilibrium Order

In order to check the accuracy of our algorithm, we have calculated the equilibrium order parameter of a collection of spherocylinders in the presence of an external field. The Helmholtz free energy density for a collection of spherocylinders can be written as:

$$\mathscr{F} = \beta F/N = \log(\Lambda^3 \rho) - 1 + \sigma[f(\underline{\Omega})] + \sum_{n=2}^{\infty} \frac{B_n \rho^{n-1}}{n-1} - \frac{\lambda}{2} S \quad (43)$$

where $\beta = 1/k_B T$, $N$ is the number of spherocylinders, $\rho = N/V$ is the number density and $\Lambda$ is the de Broglie wavelength, $\lambda$ is the strength of the applied external field and $S$ is the uniaxial nematic order parameter. The entropy of mixing term, $\sigma[f(\underline{\Omega})]$, is a functional of the one particle orientational distribution function of a spherocylinder whose orientation is described by $\underline{\Omega}$. For spherocylinders we have used the trial function:

$$f(\theta) = \frac{\sqrt{h}}{2\pi \text{Da}(\sqrt{h})} \exp[-h \sin^2 \theta] \quad (44)$$

where $\theta$ is the angle between the long axis of a spherocylinder and the nematic director/direction of the applied external field and Da(...) is the Dawson integral[28]. The parameter $h$ is related to the nematic order parameter $S = \langle P_2(\cos \theta) \rangle$ by:

$$S = \frac{1}{4}\left( \frac{3}{\sqrt{h}\text{Da}(\sqrt{h})} - \frac{3}{h} - 2 \right). \quad (45)$$

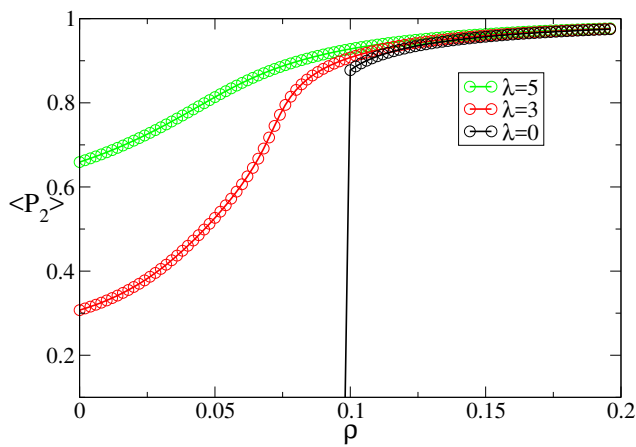

**Fig. 6** The variation of the nematic order parameter $S = \langle P_2 \rangle$ with density $\rho$ for various external field strengths.

The trial function can be used to calculate the entropy of mixing-like term as:

$$\sigma[f(\underline{\Omega})] = \int f(\underline{\Omega})\log[4\pi f(\underline{\Omega})]d\underline{\Omega},$$

$$= \log\left[\frac{4\pi\sqrt{h}}{\mathrm{Da}(\sqrt{h})}\right] + \frac{\sqrt{h}}{2\mathrm{Da}(\sqrt{h})} - h - \frac{1}{2} \quad (46)$$

The virial coefficients can be calculated from:

$$B_n = \frac{1-n}{n!V}\int\ldots\int f(\underline{\Omega}_1)\ldots f(\underline{\Omega}_n)V_n d\underline{\Omega}_1\ldots d\underline{\Omega}_n d\underline{r}_1\ldots d\underline{r}_n \quad (47)$$

with $V$ the volume, $\underline{r}_i$ and $\underline{\Omega}_i$ the position and orientation of particle $i$. $V_n$ is given by:

$$V_n = \sum_{\mathscr{S}_n} \Pi_{i<j}^n f_{ij} \quad (48)$$

$\mathscr{S}_n$ denotes that the sum is taken over all star integrals with $n$ points and $f_{ij}$ represents the Mayer f-bond between particles $i$ and $j$. For hard interactions, the Mayer f-bond is $-1$ if the particles overlap and 0 otherwise. The virial coefficients can be calculated as functions of $\alpha$ using a 'hit and miss' MC scheme. Once the virial coefficients are known, we have the Helmholtz free energy as a function of $h$, $\rho$ and $\lambda$. For fixed values of $\rho$ and $\lambda$, we minimise over $\alpha$ which gives the dependence of the nematic order parameter $S(\rho,\lambda)$. In figure 6, we show plots of $S$ vs $\rho$ for several values of the field strength $\lambda$.